\documentclass[aps,pra,reprint,amsmath,amssymb,showpacs,nofootinbib,floatfix]{revtex4-1}

\usepackage[colorlinks,citecolor=blue,urlcolor=blue,linkcolor=blue,hyperindex,breaklinks]{hyperref}

\usepackage{natbib}
\setcitestyle{authoryear,round}

\begin{document}

\title{Niels Bohr as Philosopher of Experiment: Does Decoherence Theory Challenge Bohr's Doctrine of Classical Concepts?}

\author{Kristian Camilleri} 
\address{School of Historical and Philosophical Studies, University of Melbourne, Melbourne, Victoria 3010, Australia}

\author{Maximilian Schlosshauer}
\address{Department of Physics, University of Portland, 5000 North Willamette Boulevard,
 Portland, Oregon 97203, USA}

\begin{abstract}
Niels Bohr's doctrine of the primacy of ``classical concepts'' is arguably his most criticized and misunderstood view. We present a new, careful historical analysis that makes clear that Bohr's doctrine was primarily an epistemological thesis, derived from his understanding of the functional role of experiment. A hitherto largely overlooked disagreement between Bohr and Heisenberg about the movability of the ``cut'' between measuring apparatus and observed quantum system supports the view that, for Bohr, such a cut did not originate in dynamical (ontological) considerations, but rather in functional (epistemological) considerations. As such, both the motivation and the target of Bohr's doctrine of classical concepts are of a fundamentally different nature than what is understood as the dynamical problem of the quantum-to-classical transition. Our analysis suggests that, contrary to claims often found in the literature, Bohr's doctrine is not, and cannot be, at odds with proposed solutions to the dynamical problem of the quantum-classical transition that were pursued by several of Bohr's followers and culminated in the development of decoherence theory.\\[-.2cm]

\noindent Journal reference: \emph{Stud.\ Hist.\ Phil.\ Mod.\ Phys.\ }\textbf{49}, 73--83 (2015)
\end{abstract}

\maketitle

\section{Introduction}

In spite of the attention Bohr's writings have received over the last three decades, scholarly opinion on how we should understand his philosophy remains divided \citep{Folse:1985:kc,Honner:1987:kc,Murdoch:1987:kc,Faye:1991:kc,Favrholdt:1992:kc,Faye:1994:kc,Plotnitsky:1994:kc,Brock:2003:kc,Plotnitsky:2006:kc,Katsumori:2011:kc}. Much confusion still reigns over how we should understand Bohr's repeated insistence that we \emph{must} use classical concepts. This situation is all the more lamentable, given that, as Don Howard has rightly noted, ``the doctrine of classical concepts turns out to be more fundamental to Bohr's philosophy of physics than are better-known doctrines, like complementarity'' \citep[p.~202]{Howard:1994:lm}. Scholars have long pondered over precisely why Bohr felt that classical concepts should play such a primary role in quantum physics. In perhaps his most frequently quoted account of the doctrine, in his contribution to the 1949 Einstein \emph{Festschrift}, Bohr declared:
\begin{quote}
  It is decisive to recognize that, \emph{however far the phenomena transcend the scope of classical physical explanation, the account of all evidence must be expressed in classical terms.}  The argument is simply that by the word ``experiment'' we refer to a situation where we can tell others what we have done and what we have learned and that, therefore, the account of the experimental arrangement and of the results of the observations must be expressed in unambiguous language with suitable application of the terminology of classical physics \citep[p.~209]{Bohr:1949:mz}.
\end{quote}
One can find this view, or at least anticipations of it, in Bohr's writings during the 1920s, but by the 1930s it came to occupy a central place in Bohr's epistemological reflections on quantum mechanics. Indeed Bohr was remarkably categorical about this point. As he was to put it in a lecture in the early 1930s: ``The unambiguous interpretation of any measurement \emph{must} be essentially framed in terms of classical physical theories, and we may say that in the sense the language of Newton and Maxwell will remain the language of physics for all time'' \citep[p.~692, emphasis added]{Bohr:1931:ii}. 

Over time a number of criticisms have been raised against these views of Bohr's. Most recently, spurred by the insights brought about the decoherence program \citep{Zeh:1970:yt,Zurek:1981:dd,Zurek:1982:tv,Zurek:2002:ii,Joos:2003:jh,Bacciagaluppi:2003:yz,Schlosshauer:2003:tv,Schlosshauer:2007:un}, a number of physicists have suggested that Bohr's musings about the primacy of classical concepts, and by extension his doctrine of an (ostensibly) fundamental quantum--classical divide, amount to little more than superfluous semantic or philosophical baggage, much of which has been discredited by recent developments. Dieter Zeh, for example, has contrasted the dynamical approach of decoherence with the ``irrationalism'' of the Copenhagen school \citep[p.~27]{Joos:2003:jh}.  Erich Joos, who attributes the origins of decoherence to a dissatisfaction with the ``orthodoxy of the Copenhagen school'' and ``the desire to achieve a better understanding of the quantum--classical relation'' \citep[p.~54]{Joos:2006:yy}, has argued that ``the message of decoherence'' is that ``we do not need to take classical notions as the starting point for physics,'' given that ``these emerge through the dynamical process of decoherence from the quantum substrate'' \citep[p.~77]{Joos:2006:yy}.

In this paper, we will take such claims and characterizations as our motivation for pursuing a careful historical and philosophical investigation of Bohr's views regarding his doctrine of classical concepts and the problem of the quantum--classical relationship. We will also analyze how these views relate to what we will call the ``dynamical'' approaches to the problem of the quantum-to-classical transition, approaches that include the theory of decoherence. As we shall see, decoherence is only the last step in a long line of attempts to undergird (or supplant) Bohr's doctrines by an explicit dynamical and physical account. Such approaches were already pursued by a number of Bohr's followers---notably Weizs\"acker and Rosenfeld---in the 1960s, who, far from seeing it as an invalidation of Bohr's basic insight, regarded it as providing a justification of his views.

In this paper, we raise and address two central questions. The first question is why, and in what sense, Bohr believed that classical concepts were indispensable in the description of experiments. Given the large degree of scholarly dispute and confusion about the exact meaning of Bohr's writings and his views, this requires that we pay careful attention to Bohr's texts. Here we echo Don Howard's call ``to return to Bohr's own words, filtered through no preconceived dogmas'' \citep[p.~201]{Howard:1994:lm}. In particular, we need to disentangle Bohr's views from those of his contemporaries who professed to speak on his behalf. Much of the confusion over Bohr's philosophy has resulted from a mistaken tendency to assume that Bohr's views formed the central plank in a unified and widely shared viewpoint commonly known as the ``Copenhagen interpretation.'' Yet extensive historical scholarship over the past thirty years has challenged, if not seriously undermined, the notion that any such consensus among the founders of quantum mechanics ever existed.\footnote{As Catherine Chevalley points out, ``what makes Bohr so difficult to read is the fact that his views were identified with the so-called \emph{`Copenhagen Interpretation of Quantum Mechanics,'} when such a thing emerged as a frame for philosophical discussion only in the mid-1950s'' \citep[p.~59]{Chevalley:1999:aa}. We must be clear that the term ``Copenhagen interpretation,'' as it is commonly used, refers to a range of different physical and philosophical perspectives that emerged in the decades following the establishment of quantum mechanics in the late 1920s. As Jammer points out in his \emph{Philosophy of Quantum Mechanics}: ``The Copenhagen interpretation is not a single, clear-cut, unambiguously defined set of ideas but rather a common denominator for a variety of related viewpoints. Nor is it necessarily linked with a specific philosophical or ideological position'' \citep[p.~87]{Jammer:1974:pq}. Indeed the very idea of a unitary interpretation only seems to have emerged in the 1950s in the context of the challenge of Soviet Marxist critique of quantum mechanics, and in the defense of Bohr's views, albeit from different epistemological standpoints, by Heisenberg and Rosenfeld \citep{Chevalley:1999:aa,Howard:2004:mh,Camilleri:2009:az}.}

Further complicating matters is the notoriously vague and imprecise use of the term ``classical'' in much of the literature. This term is frequently employed to refer variously to concepts, dynamical properties, phenomena, laws, or theories, without regard for the subtle but important distinctions. While Bohr often left it to his readers to decipher the precise meaning of ambiguous phrases such as ``classical description,'' in his more deliberate moments he did take care to distinguish between the use of \emph{classical concepts} (such as position and momentum) and \emph{classical dynamical theories}. In his reply to the EPR paper, for example, Bohr emphasized the necessity of using ``classical \emph{concepts} in the interpretation of all proper measurements, even though the classical \emph{theories} do not suffice in accounting for the new types of regularities with which we are concerned in atomic physics'' \citep[p.~701, emphasis added]{Bohr:1935:re}. Bohr was also careful to distinguish between our use of classical terminology and the dynamical properties of quantum objects. As he recognized, objects like electrons simply do not possess ``such inherent attributes as the idealizations of classical physics would ascribe to the object'' \citep[p.~293]{Bohr:1937:km}. Yet, Bohr repeatedly emphasized that we are simply forced to use the conceptual vocabulary of classical physics, albeit within certain limits of applicability, in describing experiments on quantum objects. Put simply, whenever we speak of an indeterminacy of an electron's \emph{position} or \emph{momentum}, we invariably fall back on the use of classical concepts. It was in this sense that Bohr used expressions such as ``the terminology of classical physics'' or the ``framework of classical physical ideas.''

The interpretation of Bohr's doctrine we present in this paper differs in many crucial respects from those that can be found in the extensive literature on Bohr. There is now consensus among Bohr scholars that his doctrine of classical concepts should be understood epistemologically. However, there is still widespread disagreement on what epistemological position Bohr held. Much of the recent literature has attempted to make sense of Bohr's views either by situating them in the context of a particular philosophical tradition, such as positivism or Kantianism, or alternatively by trying to reconstruct from Bohr's writings, a position vis-\`a-vis the contemporary realism debates.\footnote{The contributions by Favrholdt, Fay, Folse, Krips, McKinnon in the 1994 volume on Bohr all focus on the extent to which Bohr's views depart from a realist interpretation of the theory of quantum mechanics  (\citealp{Faye:1994:kc}; see also \citealp{Faye:1991:kc,Folse:1985:kc}). Murdoch, for example, has construed Bohr's disagreement with Einstein fundamentally as a debate about the realist interpretation of quantum
mechanics \citep[p.~236]{Murdoch:1987:kc}. There have been a number of efforts to draw comparisons between Bohr's views and Kantian epistemology (\citealp{Bitbol:2013:kc}; \citealp{Cuffaro:2010:kc}; \citealp{Kaiser:1992:kc}; \citealp[pp.~229--231]{Murdoch:1987:kc}; \citealp[pp.~217--221]{Folse:1985:kc}; \citealp{Honner:1982:kc}). More recently there have been renewed efforts to make sense of Bohr's writings by situating them in the Helmholtzian tradition of theoretical physics, or to read them through the lens of philosophical traditions such as hermeneutics and deconstruction \citep{Brock:2003:kc,Plotnitsky:1994:kc,Katsumori:2011:kc}.} As Henry Folse rightly points out, while it is true that Bohr's ``description of phenomenal objects has a certain Kant-like appearance,'' such an appearance is deceptive, given that complementarity has nothing to do with ``how experiences phenomena arise in the subject's consciousness'' \citep[p.~219]{Folse:1985:kc}. If we are to understand what was distinctive about Bohr's view, we cannot simply say it was grounded in an epistemological view of the primacy of classical language---rather we must ask what Bohr saw as the fundamental ``task of epistemology.'' While the attempts to characterize Bohr's view with relation to different strands of realism and antirealism have led to many important insights, and have happily led to a far more nuanced view of his philosophy than the positivist image that prevailed in the 1960s, such attempts have often inadvertently obscured Bohr's ``epistemological lesson.'' In responding to the challenge of the EPR paper, Bohr was, of course, forced to confront issues concerning the ``completeness of quantum mechanics,'' but his doctrine of classical concepts, as we stress below, was not motivated by the problem of how to interpret the quantum-mechanical formalism. Bohr's primary concern was to articulate an epistemology of experiment, not an epistemology of quantum theory. Here we adhere to Collingwood's dictum that ``we can understand a text only when we have understood the question to which it is an answer.''

This brings us to a second question, which forms another focus of this paper: What is the relationship of Bohr's doctrine of classical concepts, and in particularly his understanding of the classical description of experiment, to the problem of the quantum-to-classical transition, i.e., the problem of the ``crossings'' of the quantum--classical boundary? In this paper, we argue that much of the confusion surrounding this question stems from the failure to properly distinguish between Bohr's \emph{epistemological} thesis concerning a functional description of \emph{experiment}, and the efforts to provide a \emph{dynamical} explanation of the emergence of classicality from quantum \emph{theory}. While Bohr offered some remarks on the latter through his oblique references to the ``heaviness'' of the apparatus, his main preoccupation was with the former. A deeper understanding of this difference between Bohr's epistemological reflections on experiment and the transition between quantum and classical dynamics turns out to be crucial to overcoming many of the persistent difficulties in making sense of Bohr's views. 

In addressing these questions, this paper is divided into two main parts. The first part provides a reading of Bohr's understanding of the epistemological primacy of a classical description, based on a functional, rather than a structural, understanding of what an experiment is. While much of the literature on Bohr has focused on the use of classical concepts in any complementary description of the quantum \emph{object}, in this paper we instead devote much of our attention to his contention that the \emph{experimental apparatus} must be described in terms of the concepts of classical physics. Here the work of Peter Kroes \citep{Kroes:1994:ps,Kroes:2003:zz} turns out to be particularly instructive in bringing to light aspects of Bohr's epistemology that have otherwise been ignored. A deeper insight into Bohr's view of experiment is revealed through a close examination of his important, but often overlooked, disagreement with Heisenberg in the 1930s regarding how to understand the dividing line, or ``cut,'' between quantum object and instrument. A close reading of their correspondence, unpublished manuscripts, and published writings during this period conclusively dispels the myth that Heisenberg's articulation of the ``cut'' argument can be taken as a faithful representation of Bohr's own position. This provides further evidence of the extent to which Bohr's contemporaries, in pursuing their own philosophical agendas, often diverged from his basic insights.

The second part of this paper examines the history of attempts to come to grips with the dynamical problem of the quantum-to-classical transition. As we will show, a number of Bohr's followers attempted to provide a \emph{dynamical} reformulation of Bohr's general views after the Second World War. These attempts typically took the form of investigating the thermodynamic conditions under which interference terms would effectively vanish at the macroscopic scale. While most of these attempts ended in failure, they did not, as we shall argue, signify a departure from Bohr's general epistemological viewpoint, and they did yield some surprising insights. Perhaps most notably, Heisenberg took the crucial step of emphasizing the importance of the openness of quantum systems in his brief remarks on the problem of the quantum-to-classical transition in the 1950s. Yet it was only with the deeper understanding, brought about by decoherence theory, of the role of entanglement with the environment that the dynamical problem of the quantum-to-classical transition could be properly addressed.

This paper does not aim to mount a sustained defense of Bohr's view. Rather, our aim is to offer an interpretation of Bohr, based on a close reading of his primary texts and reconstructive analysis, that reveals the extent to which his basic insights were consistent with attempts to find a dynamical solution to the problem of the quantum-to-classical transition, which culminated with the development of decoherence theory. Though we do not aim to defend Bohr, it will become apparent that many of the standard criticisms of his viewpoint are based on a failure to understand him on his own terms. It is our hope that this investigation will establish a more refined understanding of Bohr's views, and that it will raise awareness and appreciation of the complex connections between Bohr's views, the views of his followers, and the contributions made by dynamical approaches such as decoherence.

\section{Bohr's Doctrine of Classical Concepts}

\subsection{Niels Bohr as Philosopher of Experiment}

Bohr's doctrine of the indispensability of classical concepts made a deep and lasting impression on a younger generation of physicists who at various times worked closely with Bohr in Copenhagen. Wolfgang Pauli, Werner Heisenberg, L\'eon Rosenfeld, and Aage Petersen all saw as Bohr's epistemological reflections on quantum mechanics as containing a deep insight into the nature of human knowledge. At the same time, Bohr's unwavering commitment to his view of classical concepts was an enduring source of puzzlement for many of his contemporaries. In a letter to Bohr on 13 October 1935, Schr\"odinger pressed Bohr to clarify the reasons for his view:
\begin{quote}
You have repeatedly expressed your definite conviction that measurements must be described in terms of classical concepts \dots. There must be clear and definite reasons which cause you repeatedly to declare that we \emph{must} interpret observations in classical terms, according to their very nature. \dots\ It must be among your firmest convictions---and I cannot understand what it is based upon. \cite[p.~508, emphasis in the original]{Bohr:1996:mn}.
\end{quote}
In his reply to Schr\"odinger on 26 October 1935, Bohr offered the following remarks: 
\begin{quote}
My emphasis of the point that the classical description of experiments is unavoidable amounts merely to \emph{the seemingly obvious fact that the description of any measuring arrangements must, in an essential manner, involve the arrangement of the instruments in space and their functioning in time, if we shall be able to state anything at all about the phenomena} \cite[p.~511, emphasis in the original]{Bohr:1996:mn}.
\end{quote}
Bohr's reply to Schr\"odinger appears, at least on the surface, to shed very little light on the matter. As we shall see, however, it contains an important clue to understanding Bohr's views. Here it is crucial to note Bohr's emphasis on the classical description of the experiment as a condition of the possibility of acquiring empirical knowledge of the phenomena under investigation. In Bohr's view, if a macroscopic system is to serve as a measuring instrument, it must admit of a classical description.

In order to understand why Bohr was convinced of this point, we must first recognize that from the late 1920s on, Bohr understood that the idea of a classical concept of an isolated ``object'' that has a well-defined ``state'' and that interacts with a measuring instrument is rendered problematic in quantum mechanics. Bohr recognized that this posed something of a paradox for the concept of observation. On the one hand, in order to observe something about an electron---say, its momentum---we must \emph{assume} that the electron possesses an independent dynamical state (momentum), which is in principle distinguishable from the state of the instrument with which it interacts. On the other hand, such an interaction, if treated quantum-mechanically, destroys the separability of the object and the instrument, since the resulting entanglement between the two partners means that they must be described by a single composite nonseparable quantum state. The impossibility of ``separating the behaviour of the objects from their interaction with the measuring instruments'' in quantum mechanics ``implies an ambiguity in assigning conventional attributes to atomic objects'' \citep[p.~317]{Bohr:1948:um}. A quantum-mechanical treatment of the observational interaction would paradoxically make the very distinction between object and instrument ambiguous. But such a distinction is a necessary condition for empirical inquiry. After all, an experiment is carried out precisely to reveal information about the ``autonomous behavior of a physical object'' \cite[p.~290]{Bohr:1937:km}. To speak of an interaction between two separate systems---an object and measuring instrument---is to speak in terms of classical physics. This basic point is repeated time and time again in Bohr's writings throughout the 1940s and 1950s.

For Bohr, this point lay at the heart of the epistemological paradox of quantum mechanics. He regarded the condition of isolation to be a simple logical demand, because, without such a presupposition, an electron cannot be an ``object'' of empirical knowledge at all. The crucial point, as Bohr explained at the 1936 ``Unity of Science'' congress, is that contrary to the situation in classical physics, in quantum mechanics ``it is no longer possible sharply to distinguish between the autonomous behavior of a physical object and its inevitable interaction with other bodies serving as measuring instruments,'' yet it lies in ``the very nature of the concept of observation itself'' that we can draw such a distinction \cite[p.~290]{Bohr:1937:km}. Bohr elaborated on this point at some length from his 1937 lecture ``Natural Philosophy and Human Cultures'':
\begin{quote}
We are faced here with an epistemological problem quite new in natural philosophy, where all description of experiences so far has been based on the assumption, already inherent in the ordinary conventions of language, that it is possible to distinguish sharply between the behaviour of objects and the means of observation. This assumption is not only fully justified by everyday experience, but \emph{even constitutes the whole basis of classical physics} \dots.  [In light of this situation] we are, therefore, forced to examine more closely the question of \emph{what kind of knowledge can be obtained concerning objects}. In this respect, we must \dots\ realize that \emph{the aim of every physical experiment}---to gain knowledge under reproducible and communicable conditions---\emph{leaves us no choice but to use everyday concepts, perhaps refined by the terminology of classical physics, not only in accounts of the construction and manipulation of measuring instruments but also in the description of actual experimental results} 
\citep[pp.~25--6, emphasis added]{Bohr:1987:oo}.
\end{quote}
As this passage makes clear, Bohr's philosophical preoccupations were essentially of an epistemological nature. The real question for Bohr was not what kind of reality is described by quantum mechanics, but rather ``what kind of knowledge can be obtained concerning objects'' by means of experiment. Thus, for Bohr, the epistemological problem of quantum mechanics was not how we should interpret the formalism, but rather how \emph{experimental knowledge} of quantum objects is possible. To this extent, Bohr's philosophical preoccupations were fundamentally at odds with, or at least rather different from, what many philosophers of physics see as \emph{the} problem of quantum mechanics, namely, the interpretation of its formalism. To put it simply, Bohr's doctrine of classical concepts is not primarily an \emph{interpretation} of quantum mechanics (although it certainly bears on it), but rather is an attempt by Bohr to elaborate an \emph{epistemology of experiment}.

In order to bring out this point clearly, it is instructive to borrow the terminology employed by Peter Kroes in his distinction between a \emph{structural} and a \emph{functional} description of an experimental apparatus: ``The structural description represents the object [serving as the measuring instrument] as a physical system, whereas the functional description represents the object as a technological artifact'' \citep[p.~76]{Kroes:2003:zz}. While it is always possible to conceptualize a measuring instrument, such as a mercury thermometer, ``from a purely physical (structural) point of view'' as an object with certain dynamical properties and obeying physical laws, the same object can also be described ``from an intentional (functional) point of view'' as an instrument designed ``to measure a particular physical quantity'' such as temperature.  From the epistemological point of view, the functional description is more fundamental \citep[p.~75]{Kroes:2003:zz}. As Kroes explains:
\begin{quote}
Every experiment has a goal (to measure $x$ or to detect $y$, or to show phenomenon $z$, etc.) and it is in relation to this goal that every part of the experimental setup is attributed a function, as well as actions performed during the experiment. For describing and understanding an experiment, reference to functions is unavoidable. In contrast the description of the results of the outcome of an experiment (the observations, data, measurements) is free of any reference to functions at all \dots. Thus whereas experiments are described in a functional way, the description of the results of experiments and of physical reality as constructed on the basis of those results is of a structural kind. \emph{This means that the structural description of physical reality rests implicitly on a functional description of at least part of the world} \citep[p.~74, emphasis added]{Kroes:2003:zz}.
\end{quote}
This passage provides a valuable insight through which we can make good sense of why Bohr saw it as necessary to draw an epistemological distinction between the functioning of the instrument and the object under investigation. Indeed, the language Kroes employs here bears a strikingly similarity with the way Bohr often expressed himself. In his discussion of the interaction of object and instrument, Bohr always refers to the aims and functions of the experiment. Here it is worth noting that for Kroes, a ``functional description also makes use of structural concepts; it makes reference to the structural properties of the object [serving as the measuring instrument], not only in describing but also in explaining the design features of the object'' \citep[p.~76]{Kroes:2003:zz}. 

Bohr's central insight was that if a measuring instrument is to serve its \emph{purpose of furnishing us with knowledge of an object}, it must be described classically. Of course, it is always possible to represent the experimental apparatus from a purely \emph{structural} point of view, without any reference to its function, as a \emph{quantum-mechanical system}. However, any functional description of the experimental apparatus, in which it is treated as a means to an end and not merely as a dynamical system, must make use of the concepts of a classical physics. This is true even when we measure quantum properties, such as spin. In the Stern--Gerlach experiment, for example, the experimental apparatus and the magnetic field must be treated classically for the purpose of performing the experiment. Put simply, any functional description of the experimental apparatus must be a \emph{classical} description. In Bohr's view, ``all unambiguous information concerning atomic objects is derived from permanent marks---such as a spot on a photographic plate, caused by the impact of an electron---left on the bodies which define the experimental conditions'' \citep[p.~310]{Bohr:1958:mj}. In his 1958 lecture ``On Atoms and Human Knowledge,'' Bohr expanded on this point:
\begin{quote}
In the analysis of single atomic particles, this is made possible by irreversible amplification effects---such as a spot on a photographic plate left by the impact of an electron, or an electric discharge created in a counter device---and the observations concern only \emph{where and when the particle is registered on the plate or its energy on arrival with the counter}. Of course, this information presupposes knowledge of the position of the photographic plate relative to other parts of the experimental arrangement, such as regulating diaphragms and shutters defining space-time coordination or electrified and magnetized bodies which determine the external force fields acting on the particle and permit energy measurements. The experimental conditions can be varied in many ways, but the point is that in each case we must be able to communicate to others what we have done and what we have learned, and that therefore \emph{the functioning of the measuring instruments must be described within the framework of classical physical ideas} \citep[pp.~169--170, emphasis added]{Bohr:1958:lu}.
\end{quote}
The properties of the quantum object, such as its position, charge, spin and energy, can only be known by virtue of traces it leaves on an experimental system. It must be possible to interpret the results of the experiment in such a way that the visible and permanent traces on the apparatus can be dynamically explained as having been caused by the object under investigation. If this were not the case---if no such causal inferences were possible---then such systems could not provide us with knowledge of the dynamical properties of the quantum object. To this end, we must presuppose a causal chain of events triggered by the object itself, through the apparatus, finally registering at the macroscopic scale if the measuring apparatus is to serve its purpose. Or, to put it another way, it must be possible to trace this ``sequence of cause and effect'' back from the observation of a spot on a photographic plate or an electric discharge in a counter device, to the object itself. In this sense, ``the concept of causality underlies the very interpretation of each result of experiment'' insofar as it forms the basis of any functional description \citep[p.~293]{Bohr:1937:km}.

\subsection{The ``Cut'' between Object and Apparatus: Bohr's Disagreement with Heisenberg}

The reading of Bohr presented above, it should be stressed, does not presuppose any particular ontological division of the world into quantum and classical realms. There is, for Bohr, no fundamental ontological difference between the quantum object and the experimental apparatus we use to investigate the world. Much of the confusion surrounding Bohr's views stems from a tendency to read him as having ``split the physical world into a quantum microcosm and a classical macrocosm''
\citep[p.~102]{Osnaghi:2009:aa}. Yet, Bohr's discrimination ``between those parts of the physical system \dots\ which are to be treated as measuring instruments and those which constitute the objects under investigation'' was determined by functional-structural (epistemological) considerations and not by dynamical (ontological) considerations \citep[p.~701]{Bohr:1935:re}.

However, a deeper understanding of Bohr's epistemological viewpoint requires us to disentangle it from a viewpoint that has often been attributed to him. Bohr did not, as is commonly assumed, see the object–-instrument divide as entailing a quantum--classical divide, at least not in the sense of different realms of dynamical laws. This view of Bohr has persisted, in large part, because it was a view attributed to Bohr by many of his contemporaries, and even by some of his closest collaborators. The chief offender here is Heisenberg, whose discussion of the ``cut'' between the object and instrument in the 1930s has shaped the contemporary reading of Bohr. As we shall see, however, Bohr never endorsed Heisenberg's view of the cut, and indeed raised certain objections to Heisenberg in their private correspondence. According to this commonly held view, which Don Howard has labeled the \emph{coincidence interpretation}, ``the experimental apparatus is to be given a classical description in its entirety, while the object is described by means of quantum mechanics'' \cite[p.~203]{Howard:1994:lm}. Yet, contrary to what some commentators have assumed, Bohr never endorsed such a view.\footnote{Although we agree with Howard that Bohr did not endorse the coincidence interpretation, our reading differs from Howard's ``reconstruction'' of Bohr's doctrine of classical concepts. According to Howard, Bohr's mutually exclusive experimental arrangements may be identified with the choice of effective mixed states (ignorance-interpretable mixtures; see \citealp{Espagnat:1988:cf}) conditioned on the particular measurement setup \citep{Howard:1994:lm}. Howard emphasized that ``Bohr required a classical description of \emph{some}, but not necessarily \emph{all}, features of the instrument and more surprisingly, perhaps, a classical description of some features of the observed object as well'' \cite[p.~203]{Howard:1994:lm}.}  Erhard Scheibe, in his otherwise penetrating analysis of Bohr's views, attributes to Bohr the coincidence interpretation in his discussion of the use of mutually exclusive experimental arrangements in measuring the position and momentum of a particle. In Scheibe's view, Bohr held that in such cases, the ``part of the experimental arrangements that is initially described in classical terms is converted into the object and thus \emph{must be described in quantum-theoretical terms}, whereas the requirement of classical description is applied to different parts of the experimental arrangement'' \citep[p.~48, emphasis added]{Scheibe:1973:kc}. This view, however, rests on a misreading of Bohr.

Put simply, Bohr never described the measurement interaction between the quantum object and the apparatus as an interaction between an object described by a wave function in configuration space and a measuring instrument described by classical dynamical laws in ordinary three-dimensional space. As Howard rightly notes, this reading of Bohr is deeply problematic insofar as it ``introduces a dualism into our ontology''; it also raises the further problem of how the two systems ``described by fundamentally different theories'' interact with each other, given that such an interaction does not appear to be subsumed under either classical or quantum theory \cite[p.~211]{Howard:1994:lm}. Bohr insisted that the concepts of classical physics must be used on \emph{both} sides of the diving line. This is because, as Bohr emphasized, ``a measurement can mean nothing else than the unambiguous comparison of some property of the object under investigation with a corresponding property of another system, serving as a measuring instrument'' \citep[p.~19]{Bohr:1939:ww}. Such comparisons invariably involve the use of classical concepts.

Some clarification is in order here. In Bohr's view, whereas a functional description of the \emph{experimental apparatus} must make some use of \emph{classical dynamical theories} (like Newton's or Maxwell's laws), the corresponding structural description of the experimental object can only be a partial, classical description. Here we are obliged to make use of different classical concepts in mutually exclusive experimental arrangements. As Bohr would put it, ``evidence obtained under different experimental conditions cannot be comprehended within a single picture, but must be regarded as \emph{complementary} in the sense that only the totality of the phenomena exhausts the possible information about the objects'' \cite[p.~210]{Bohr:1949:mz}. To this extent, Bohr argued that in our experimental investigation of the kinematic-dynamic behaviour of quantum objects, we are forced to ``use two different experimental arrangements, of which only one permits the unambiguous \emph{use} of the concept of position, while only the other permits the \emph{application} of the concept of momentum'' \citep[p.~293, emphasis added]{Bohr:1937:km}. To this extent, the situation in quantum mechanics brings to light the ``previously unrecognized presuppositions for an unambiguous use of our most simple concepts'' \cite[p.~290]{Bohr:1937:km}. While a more detailed investigation of complementarity is beyond the scope of this paper, it should be clear that the coincidence interpretation does not form part of Bohr's general outlook.

Having now clarified Bohr's view, we can turn our attention to Heisenberg. In a widely read article appearing in \emph{Naturwissenschaften} in 1934 \citep{Heisenberg:1952:mn}, Heisenberg gave perhaps his earliest articulation of the coincidence  interpretation. There he argued that ``in a mathematical treatment of the process, a dividing line must be drawn between, on the one hand, the apparatus which we use as an aid in putting the question and thus, in a way, treat as part of ourselves, and on the other hand, the physical systems we wish to investigate'' \cite[p.~15]{Heisenberg:1952:mn}. Heisenberg went on to construe this dividing line as a cut [\emph{Schnitt}] ``between the measuring apparatus of the observer which is described in \emph{classical concepts}, and the object under observation, whose behavior is represented by a \emph{wave function}'' \cite[p.~15, emphasis added]{Heisenberg:1952:mn}. This view was reiterated in a lecture in November 1935 in Vienna, in which Heisenberg argued that where we chose to place the diving line between the quantum and classical description ``is immediately defined by the nature of the problem but it obviously \emph{signifies no discontinuity of the physical process}. For this reason there must, within certain limits, exist complete freedom in choosing the position of the dividing line'' \citep[p.~49, emphasis added]{Heisenberg:1952:mk}.

Perhaps the most developed version of the ``cut'' argument can be found in Heisenberg's unpublished manuscript, entitled ``Ist eine deterministische Er\-g\"an\-zung der Quantenmechanik m\"oglich?'' (``Is a deterministic completion of quantum mechanics possible?'') and written some time in June 1935. Heisenberg wrote the paper in response to the Einstein--Podolsky--Rosen paper and the recent criticisms from physicists like Schr\"odinger and Laue, with the intention of submitting it for publication in the journal \emph{Naturwissenschaften}. After learning, however, that Bohr was drafting his own reply to the EPR paper, he decided against publication. He sent a copy of the manuscript to Pauli on 2 July 1935 \cite[pp.~409--18, item 414]{Pauli:1985:za}, and enclosed a copy in a letter to Bohr on 10 August, for comment. In the manuscript, Heisenberg emphasized that in any measurement scenario, the position of the ``cut can be shifted arbitrarily far in the direction of the observer in the region that can otherwise be described according to the laws of classical physics,'' but ``the cut cannot be shifted arbitrarily in the direction of the atomic system'' described, as it is, by the Schr\"odinger wave equation in configuration space \cite[p.~414]{Heisenberg:1985:zq}. Heisenberg acknowledged that strictly speaking, the laws of quantum mechanics are applicable to \emph{all} systems (including the measuring instrument), but that ``it is precisely the arbitrariness in the choice of the location of the cut that is decisive for the application of quantum mechanics'' \cite[p.~416]{Heisenberg:1985:zq}.\footnote{Elise Crull and Guido Bacciagaluppi have recently translated Heisenberg's unpublished paper into English and have made a detailed critical analysis of the argument \citep{Crull:2011:aa}. Crull notes that the structure of Heisenberg's argument is founded on two premises: first, that the location of the cut is arbitrary (i.e., that the predictions of quantum mechanics are independent of the position of the cut), and secondly, that the statistical element of quantum mechanics can only enter at the location of the cut. The paper offers a defense of both premises, from which Heisenberg draws the conclusion that a deterministic completion of quantum mechanics is impossible. However, as Crull points out, Heisenberg's formal treatment of the measurement interaction contains a flaw, in that the object and the measuring apparatus are represented by separable wave functions immediately after the interaction between them has taken place. Here Heisenberg fails to recognize that according to quantum mechanics, they will remain in an entangled state after the interaction has taken place.}

Heisenberg's view of the mobility of the quantum--classical divide is often taken to be a central planck of Bohr's view of quantum mechanics. According to Wojciech Zurek: ``The key feature of the Copenhagen Interpretation is the dividing line between quantum and classical. Bohr emphasized that the border must be mobile'' \citep{Zurek:1991:vv}. However, such pronouncements are misleading, and by September 1935, it is evident that there were crucial, and often-overlooked, differences between Bohr and Heisenberg. While Bohr never openly or publicly criticized Heisenberg's view of ``cut,'' a close reading of their correspondence, written in August and September of 1935, brings to light their hidden disagreement  \citep[Heisenberg to Bohr 10 August 1935, Bohr to Heisenberg 10 September 1935, Bohr to Heisenberg 15 September 1935, Heisenberg to Bohr 29 September 1935]{AHQP:1986:po}. While in his reply to the EPR paper (which he submitted to \emph{Physical Review} in July 1935), Bohr did refer to the ``free choice'' of where we choose to place the object--apparatus divide, he never expressed this divide in terms of the distinction between the quantum and classical mechanics. Writing to Heisenberg after reading Heisenberg's manuscript, Bohr admitted he found himself unable to agree with the line of argument contained in the paper:
\begin{quote}
I have experienced difficulties by trying to understand more clearly the argumentation in your article. For I am not quite sure that I fully understand the importance you attach to the freedom of shifting the cut between the object and the measuring apparatus. Any well-defined quantum-mechanical problem must be concerned with a certain classically described experimental setting, and if one changes the kind or use of the measuring instruments, and thus the setting, the phenomenon will always change completely. \emph{I therefore believe that for a given experimental setting the cut is determined by the nature of the problem} \dots. Moreover, I do not fully understand how the mentioned narrow possibility of shifting the cut at any given experimental setting can be used in connection with the particular problem of measurement, which you proceed to deal with
\citep[Bohr to Heisenberg, 15 September 1935]{AHQP:1986:po}.\footnote{We would like to thank Helge Kragh for translating Bohr's letter from the original Danish.}
\end{quote}
As the letter to Heisenberg makes clear, and contrary to what John Bell and many other commentators have assumed, Bohr evidently did \emph{not} ``take great satisfaction'' in ``the shiftiness of the division between `quantum system' and `classical apparatus'\,'' \citep[p.~189]{Bell:1987:pw}. In Bohr's view, once the aims of the experiment had been decided upon and the experimental apparatus has been set up accordingly, the ``cut'' was effectively fixed. It could not be moved around arbitrarily. To this extent, Bohr argued that the cut corresponds to something ``objective'' in the sense that the object--instrument distinction was essentially fixed by the \emph{functional-epistemological} considerations dictated by the choice of the particular experimental arrangement.

In a letter replying to Bohr in September 1935, Heisenberg simply reiterated his view that without the presupposition of a moving cut, one would have to conclude that there exist ``two categories of physical systems---classical and quantum-mechanical ones'' \cite[Heisenberg to Bohr 29 September 1935]{AHQP:1986:po}. Bohr's criticisms, which were guided by epistemological considerations, appear to have had little effect. Heisenberg presented the same argument we find in his unpublished manuscript, albeit in a somewhat abbreviated form, in his Vienna lecture in November 1935. Here it seems Heisenberg simply assumed that the empirical object could be described as a wave function in configuration space. Yet for Bohr, ``the appropriate physical interpretation of the symbolic quantum-mechanical formalism amounts only to predictions, of determinate or statistical character, pertaining to individual phenomena appearing under conditions defined by classical physical concepts'' \cite[p.~238]{Bohr:1949:mz}. In order to describe the particular \emph{empirical phenomena} that reveal themselves under experimental conditions, we must use the concepts of classical physics. 

This fundamental disagreement between Bohr and Heisenberg was never resolved. In his 1955--6 Gifford lectures, Heisenberg openly acknowledged that in contrast with his own view of the movability of the cut, ``Bohr has emphasized that it is more realistic to state that the division into the object and rest of the world is not arbitrary'' but rather is determined by the very nature of the experiment \cite[p.~24]{Heisenberg:1989:zb}. This point is repeated in a letter to Patrick Heelan in 1975, in which Heisenberg again drew attention to the fact that he and Bohr had never reached agreement about ``whether the cut between that part of the experiment which should be described in classical terms and the other quantum-theoretical part had a well defined position or not'' \cite[p.~137]{Heelan:1975:kk}. As he recalled: ``I argued that a cut could be moved around to some extent while Bohr preferred to think that the position is uniquely defined in every experiment'' \cite[p.~137]{Heelan:1975:kk}. Weizs\"acker also provides a similar account of this difference of opinion between Bohr and Heisenberg, which first came to light in the 1930s \cite[p.~283]{Weizsacker:1987:ym}. A historical understanding of this disagreement provides a crucial insight into Bohr's doctrine of classical concepts.

\section{The Problem of the Quantum-to-Classical Transition}

\subsection{The Equivalence of Quantum and Classical Descriptions} 

Bohr's doctrine of classical concepts, as we have argued in the previous section, is an epistemological thesis based on a functional understanding of experiment. Yet, as many commentators have pointed out, while Bohr insisted that we must employ the concepts of classical physics to describe whatever part of the system we have designated to \emph{function} as a measuring instrument, it is always \emph{possible} to give a quantum-mechanical description of the apparatus in its entirety. Bohr acknowledged that measuring instruments, like all systems, macroscopic or microscopic, are strictly speaking subject to the laws of quantum mechanics. As he explained at the Warsaw conference in 1938:
\begin{quote}
In the system to which the quantum mechanical formalism is to be applied, it is of course possible to include any intermediate auxiliary agency employed in the measuring process. Since, however, all those properties of such agencies which, according to the \emph{aim of measurements} have to be compared with the corresponding properties of the object, \emph{must be described on classical lines, their quantum mechanical treatment will for this purpose be essentially equivalent with a classical description} \citep[pp.~23--24, emphasis added]{Bohr:1939:ww}.
\end{quote}
While it is of course always \emph{possible} to describe the apparatus as a quantum-mechanical system, Bohr insists that in doing so we would forfeit a functional account of the experiment as a means of acquiring empirical knowledge. Measuring instruments, for Bohr, must admit of a classical description, otherwise they could not perform their epistemic function as measuring instruments. To this extent, any ``quantum mechanical treatment'' of a measuring instrument will, by virtue of its \emph{function} as a measuring instrument, ``be essentially equivalent with a classical description.'' However, this raises the further question of why, from a purely dynamical point of view, the quantum-mechanical treatment is essentially equivalent to a classical description. As Kroes points out, from such a structural point of view, the design and geometric configuration of the apparatus ``is just some physical property'' and ``a completely contingent feature of the object'' \cite[p.~76]{Kroes:2003:zz}.

Bohr's \emph{epistemological} explanation for \emph{why} we must use a classical description thus begs the question of what \emph{dynamical} features of a macroscopic system entitle us to neglect the ``quantum effects.'' Bohr here appears to simply assume that there exists a macroscopic ``region where the quantum-mechanical description of the process concerned is effectively equivalent with the classical description'' \citep[p.~701]{Bohr:1935:re}. Thus we are led to ask: How is it that classical physics \emph{can be} employed, at least to a very good approximation, under certain dynamical conditions (typically those corresponding to measuring scenarios)? This is a salient question, given that, strictly speaking, the world, as Bohr recognized, \emph{is} nonclassical.

Bohr never provided a sustained or satisfactory dynamical explanation for the quantum-to-classical transition, leaving only fleeting remarks scattered throughout his writings. He seems to have regarded the explanation as trivial, and on most occasions has been content to refer to the ``massive'' nature of macroscopic bodies serving as measuring instruments. A functional account of the experiment ``is secured by the use, as measuring instruments, of rigid bodies \emph{sufficiently heavy} to allow a completely classical account of their relative positions and velocities'' \citep[p.~310, emphasis added]{Bohr:1958:mj}. One finds a similar view expressed on several occasions in Bohr's later writings. As he put it in 1958, ``all measurements thus concern bodies \emph{sufficiently heavy} to permit the quantum [effects] to be neglected in their description'' \citep[p.~170, emphasis added]{Bohr:1958:lu}. But, one might well ask, how heavy is ``sufficiently heavy''?

In his reconstruction of Bohr's views, Howard attempts to provide an answer to the question of how Bohr thought it was possible that a classical description can be ``essentially equivalent to a quantum mechanical one'' \citep[pp.~217]{Howard:1994:lm}. While Howard regards his reconstruction, which employs a formal quantum-mechanical description of experimental arrangements in terms of a ``mixtures model,'' as perfectly ``consistent with Bohr's remarks on observation and classical concepts,'' he admits there is no evidence to suggest that Bohr ever explicitly considered an ``appropriate mixtures'' model \citep[pp.~225]{Howard:1994:lm}. To this end, Howard concedes that ``one is forced to go beyond the record of Bohr's words and their meanings'' \citep[pp.~204]{Howard:1994:lm}. Many of these difficulties can be avoided, however, if we distinguish between Bohr's epistemological doctrine of classical concepts (on which he said a great deal) and his views on the quantum--classical transition (on which Bohr said very little). On the few occasions when Bohr did discuss the latter, he typically appealed to the ``heaviness'' or the ``macroscopic dimensions'' of the measuring apparatus. Yet, as Hugh Everett stressed in a letter to Aage Petersen in May 1957, this assumption was problematic for the ``Copenhagen school'':
\begin{quote}
You talk of the massiveness of macrosystems allowing one to neglect further quantum effects \dots\ but never give any justification of this flatly asserted dogma. Is it an independent postulate? It most certainly does not follow from wave mechanics \dots. In fact by the very formulation of your viewpoint you are totally incapable of any justification and \emph{must} make it an independent postulate---that macrosystems are relatively immune to quantum effects. 
(Everett to Petersen, 31 May 1957, Wheeler Papers, Series I---Box Di---Fermi Award \#1---Folder Everett, quoted in \citealp[p.~106]{Osnaghi:2009:aa}).
\end{quote}
Everett's comments here go to the heart of the matter. While he never seems to have appreciated the subtleties of Bohr's epistemological position, he did pinpoint the dynamical problem of the quantum-to-classical transition. Bohr appears to have simply \emph{assumed} that any macroscopic system serving as a measuring instrument \emph{must be} describable by means of a classical approximation---otherwise we could not rely on such system to perform experiments. It is simply the case that without such a presupposition, experimental knowledge would be rendered impossible. The task then fell to Bohr's followers to provide an adequate dynamical account of why this is so. Throughout the 1950s and 1960s, a number of Bohr's disciples, including Rosenfeld, Petersen, and Groenwald, became increasingly preoccupied with the problem of the quantum-to-classical transition. 

In their article on the reception of Everett's relative-state interpretation in Copenhagen, Osnaghi, Freitas, and Freire make the observation that ``the arguments put forward by the Copenhagen group'' often ``involved (and sometimes mixed up) two different levels of reflection'' \citep[p.~116]{Osnaghi:2009:aa}. On the one hand, the ``Copenhagen group'' embraced Bohr's ``\emph{pragmatic-transcendental} argument'' for the possibility of experimental knowledge, while on the other hand, they also attempted to give ``a \emph{physical} explanation'' of the quantum-to-classical transition for macroscopic systems \citep[p.~116]{Osnaghi:2009:aa}. This attempt---the reformulation of Bohr's \emph{epistemological} doctrine in \emph{physical} terms---assumed crucial importance in the 1950s and 1960s and generated much debate and discussion. What is perhaps remarkable about this situation is that both critics and defenders of Bohr's viewpoint saw the need to arrive at a deeper understanding of the quantum-to-classical transition.

\subsection{\label{sec:irrev-emphdyn-probl}Irreversibility and the Dynamical Problem of the Quantum-to-Classical Transition}

In the 1950s and 1960s, a number of Bohr's followers turned their attention to the dynamical problem of the quantum-to-classical transition amidst the new wave of criticisms of quantum mechanics. Commenting on recent attempts in this direction in 1965, Rosenfeld argued that ``it is understandable that in order to exhibit more directly the link between the physical concepts and their mathematical representation, a more formal rendering of Bohr's argument should be attempted'' \cite[p.~536]{Rosenfeld:1979:bb}. In spite of their philosophical differences, many of Bohr's followers, such as Weizs\"acker and Rosenfeld, pursued this kind of approach to the physics of the quantum-to-classical transition as entirely in keeping with the spirit in which Bohr had intended his doctrine of classical concepts. 

As Weizs\"acker put it at a colloquium in 1968, ``the crucial point in the Copenhagen interpretation'' is captured, ``but not very luckily expressed, in Bohr's famous statement that all experiments are to be described in classical terms'' \cite[p.~25]{Weizsacker:1971:ll}. As a devotee of Bohr, this was a view that Weizs\"acker endorsed wholeheartedly, but which he now wished to justify. ``My proposed answer is that Bohr was essentially right'' in arguing that the instrument must be classically describable, ``but that he did not know why'' \cite[p.~28]{Weizsacker:1971:ll}. The paradox at the heart of the Copenhagen interpretation for Weizs\"acker is therefore to be stated: ``Having thus accepted the falsity of classical physics, taken literally, we must ask how it can be explained as an essentially good approximation'' when describing objects at the macrolevel. He spells this out:
\begin{quote}
  This amounts to asking \emph{what physical condition must be  
  imposed on a quantum-theoretical system in order that it should 
  show the features which we describe as ``classical.''} My 
  hypothesis is that this is precisely the condition that it should    
  be suitable as a measuring instrument. If we ask what that 
  presupposes, a minimum condition seems to be that irreversible 
  processes should take place in the system. For every measurement   
  must produce a trace of what
  has happened; an event that goes completely unregistered is not a
  measurement. Irreversibility implies a description of the system in
  which some of the information that we may think of as being present
  in the system is not actually used. Hence the system is certainly
  not in a ``pure state''; we will describe it as a ``mixture.'' I am
  unable to prove mathematically that the condition of 
  irreversibility would suffice to define a classical approximation, 
  but I feel confident it is a necessary condition \cite[p.~29, 
  emphasis in original]{Weizsacker:1971:ll}.
\end{quote}
Weizs\"acker's remarks here point to a program of dynamical explanation that had been pursued since the 1950s. After the Second World War, a number of physicists had devoted themselves to investigating the thermodynamic conditions of irreversibility that would need to hold in order for a measurement to be registered macroscopically as ``classical.'' Bohr himself had hinted at the idea that ``the essential irreversibility inherent in the very concept of observation'' might provide the clue to explain the ``classicality'' of quantum systems at the macroscopic level as a thermodynamic irreversible process on a number of occasions \citep[p.~310]{Bohr:1958:mj}. ``The amplification of atomic effects,'' as he noted, ``emphasizes the irreversibility characteristic of the very concept of observation'' \citep[p.~170]{Bohr:1958:lu}. Yet, as Everett pointed out in 1956, ``there is nowhere to be found any consistent explanation of this `irreversibility' attributed to the measuring process'' (Everett's notes on Stern's letters 1956, as quoted in \citealp[p.~106]{Osnaghi:2009:aa}).

A more rigorous ``thermodynamic approach'' was developed independently by G\"unther Ludwig in the second half of the 1950s \citep[p.~488]{Jammer:1974:pq}. Ludwig attempted to explain measurement by means of a thermodynamic analysis of the irreversible amplification process triggered by a microscopic event \citep[p.~490]{Jammer:1974:pq}. As Osnaghi, Freitas, and Freire explain, many physicists saw Ludwig's approach as opening up ``the possibility of providing a rigorous foundation for Bohr's approach, thereby dispelling the misunderstandings surrounding the alleged subjectivism of the Copenhagen view'' \citep[p.~103]{Osnaghi:2009:aa}. A number of attempts to develop further Ludwig's basic program of a thermodynamic approach were pursued during the 1960s, the most elaborate of which was the work carried out by the Italian physicists Daneri, Loinger, and Prosperi \citep{Daneri:1962:om}. Their aim was to find the exact ergodicity conditions for the validity of the ergodic theorem in quantum statistical mechanics. As Jeffery Bub explained, the question the authors wished to answer was, ``How does the \emph{theory} guarantee that quantum theoretical macrostates will not exhibit interference effects?'' \citep[p.~66]{Bub:1971:ll}. The answer Daneri, Loinger, and Prosperi gave was that the physical structure of a large body implies ergodicity conditions, which in turn prevent the system from exhibiting quantum superpositions at the macroscopic scale \citep[see also][]{Ludwig:1953:oo,Ludwig:1955:oo}.

In Rosenfeld's view, this work represented a major step forward in clearing up the misunderstandings and the ``extravagant speculations'' on the measurement problem, which had arisen through ``the deficiencies in von Neumann's axiomatic treatment'' \cite[p.~537]{Rosenfeld:1979:bb}. Importantly, Rosenfeld declared that the Daneri--Loinger--Prosperi approach was ``in complete harmony with Bohr's ideas'' \cite[p.~539]{Rosenfeld:1979:bb}. Jeffery Bub, on the other hand, saw it as ``basically opposed to Bohr's ideas'' insofar as it treated classical mechanics as an ``approximation to a quantum theory of macroscopic systems (\citealp{Bub:1968:yy}; see also \citealp[p.~65]{Bub:1971:ll}). In a similar vein, Max Jammer called into question whether the approach of Daneri, Loinger, and Prosperi was ``really congenial, or at least not incompatible, with the basic tenets of the Copenhagen interpretation'' \cite[p.~493]{Jammer:1974:pq}. Yet as we have seen, the \emph{epistemological} primacy of classical physics, on which the functional description of experiment rested, was for Bohr perfectly compatible with, and indeed depended on, the view that classical mechanics was an approximation of quantum mechanics.
Those physicists closest to Bohr appreciated this point. As Weizs\"acker put it, Bohr's emphasis on ``the classical description of an instrument just meant that only so far as it was an approximation would the instrument be of use \emph{as} an instrument'' (Weizs\"acker, 1987, p. 283). There was never any suggestion for Bohr that measuring instruments could not in principle be described by quantum mechanics, or that classical dynamics was somehow more fundamental in an ontological sense than quantum mechanics. 

Yet the subtleties of Bohr's point of view were typically lost on other physicists, many of whom saw little value in Bohr's forays into epistemology. This episode is indicative of the state of confusion surrounding Bohr's doctrine of classical concepts, which persisted well into the 1960s and early 1970s. In a qualified sense, Rosenfeld was right---there was no essential conflict between Bohr's \emph{epistemological} view of the doctrine of classical concepts and the attempt to find a \emph{dynamical} solution  to the problem of the quantum-to-classical transition. 

Nevertheless, an adequate dynamical solution turned out to be far more complicated than Bohr or many of his followers had originally thought. Rosenfeld's endorsement of the ergodic solution as having provided the solution to the problem was based on his view that ``the reduction rule'' is ``essentially a thermodynamic effect, and, accordingly, only valid to the thermodynamic approximation'' (Rosenfeld to Belinfante, 24 July 1972, Rosenfeld Papers, quoted in \citealp{Osnaghi:2009:aa}, footnote 243). In hindsight, such pronouncements appear to have been overly optimistic. Eugene Wigner, who retained a key interest in foundational questions of quantum mechanics throughout the 1970s, rightly pointed out that the ergodicity conditions of Daneri--Loinger--Prosperi, which purported to explain ``the transition to a classical description of the apparatus,'' rested on ``an arbitrary step,'' which only served to ``postulate the miracle which disturbs us'' \cite[p.~65]{Wigner:1995:jm}. In a similar vein, Bub argued that the authors had simply reintroduced ``von Neumann's infinite regress all over again'' \citep[p.~70]{Bub:1971:ll}. The problem of the quantum-to-classical transition remained unsolved. On this point, Wigner and Bub were right.

\subsection{\label{sec:heis-relat-betw}The Relation between Object, Instrument, and the External World}

Looking back over the history of the foundations of quantum mechanics, it is easy to see that the crucial obstacle to an understanding of the quantum-to-classical transition was the erroneous assumption that we can treat quantum systems as isolated from the environment. There do, however, appear to be anticipations of the relevance of the environment in the 1950s and 1960s. In his contribution to the Bohr \emph{Festschrift} edited by Pauli in 1955, Heisenberg reminded his readers that a quantum-mechanical treatment of the ``interaction of the system with the measuring apparatus'' does not of itself ``as a rule lead to a definite result (e.g., the blackening of a photographic plate)'' \cite[p.~22]{Heisenberg:1955:lm}. Inclusion of further systems, such as a secondary apparatus or a human observer, will not terminate the resulting von Neumann chain \citep{vonNeumann:1955:ii} if these systems are treated as interacting quantum systems.

Heisenberg observed that this situation had prompted a number of physicists to attempt to develop a modified dynamics of wave-packet collapse, or to suggest that the conscious observer plays an essential role in the process of measurement. Yet, as Heisenberg explained, such attempts failed to grasp the real nature of the problem. In the von Neumann scheme, ``the apparatus and the system are regarded as cut off from \emph{the rest of the world} and treated as a whole according to quantum mechanics'' \cite[p.~22, emphasis added]{Heisenberg:1955:lm}. Yet the measuring apparatus is in reality never isolated from its environment: ``If the measuring device would be isolated from the rest of the world, it would be neither a measuring device nor could it be described in the terms of classical physics at all'' \cite[p.~24]{Heisenberg:1989:zb}. Heisenberg emphasized that ``the connection with the external world is one of the necessary conditions for the measuring apparatus to perform its function'' \cite[pp.~26--7]{Heisenberg:1955:lm}. Heisenberg spelled out this position in more detail in \emph{Physics and Philosophy}. Here it is worth quoting Heisenberg at some length:
\begin{quote}
  Again the obvious starting point for the physical interpretation of
  the formalism seems to be the fact that mathematical scheme of
  quantum mechanics approaches that of classical mechanics in
  dimensions which are large compared to the size of atoms. But even
  this statement must be made with some reservations. Even in large
  dimensions there are many solutions of the quantum-mechanical
  equations to which no analogous solutions can be found in classical
  physics. In these solutions the phenomenon of the ``interference of
  probabilities'' would show up \dots\ [which] does not exist in
  classical physics. Therefore, even in the limit of large dimensions
  the correlation between the mathematical symbols, the measurements,
  and the ordinary concepts is by no means trivial. In order to get at such an   unambiguous correlation one must take another feature of the problem   into account. It must be observed that the system which is treated
  by the methods of quantum mechanics is in fact a part of a much
  bigger system (eventually the whole world); it is interacting with
  this bigger system; and one must add that the microscopic properties
  of the bigger system are (at least to a large extent) unknown. This
  statement is undoubtedly a correct description of the actual
  situation \dots. The interaction with the bigger system with its
  undefined microscopic properties then introduces a new statistical
  element into the description \dots\ of the system under
  consideration. In the limiting case of the large dimensions this
  statistical element destroys the effects of the ``interference of
  probabilities'' in such a manner that the quantum-mechanical scheme
  really approaches the classical one in the limit
  \cite[pp.~121--2]{Heisenberg:1989:zb}.
\end{quote}
In this intriguing passage, Heisenberg seems to have recognized that one cannot simply appeal to the macroscopic dimensions or the mass of the measuring apparatus to explain why the apparatus is never found in a quantum superposition. Such a view marks a departure from Bohr's suggestion that the heaviness of the apparatus that renders it effectively classical. Heisenberg enlarges the system--apparatus composite to include couplings to further degrees of freedom in the environment (the ``external world''). This is a most interesting point. Of course, for practical purposes, Heisenberg admits that we often treat the quantum system and the measuring apparatus as isolated from the rest of the world. But, in Heisenberg's view, a proper account of the quantum-to-classical transition must in the end rest on ``the underlying assumption,'' which Heisenberg took to be implicit in the Copenhagen interpretation, ``that the interference terms are in the actual experiment removed by the partly undefined interactions of the measuring apparatus, with the system and with the rest of the world (in the formalism, the interaction produces a `mixture')'' \cite[p.~23]{Heisenberg:1955:lm}.

\subsection{Decoherence Theory and ``Emergent Classicality''}

Heisenberg's remarks might be read, somewhat charitably, as anticipating certain results of the decoherence program. After all, it is the hallmark of decoherence that it proceeds from the recognition that it is practically impossible to isolate the quantum system and the apparatus from the surrounding environment, and moreover that it is precisely this feature that results in the emergence of classicality. But while Heisenberg emphasizes the importance of environmental interactions, he gives no detailed account of precisely how ``the interference terms are \dots\ removed by the partly undefined interactions of the measuring apparatus, with the system and with the rest of the world'' \cite[p.~23]{Heisenberg:1955:lm}. In particular, nowhere does he explicate the specific role of entanglement between the system and the environment as the crucial point in the dynamical emergence of classicality in the system.   

While physicists such as Bohr, Heisenberg, and Schr\"odinger had recognized the nonseparability of quantum systems---i.e., entanglement---as a characteristic feature of quantum mechanics as early as the 1920s and 1930s,\footnote{Witness, for example, Schr\"odinger, who had coined the term ``entanglement'' (\emph{Verschr\"ankung} in German) in 1935 \citep{Schrodinger:1935:gs,Schrodinger:1935:jn,Schrodinger:1936:jn} and referred to it not as ``\emph{one} but rather \emph{the} characteristic trait of quantum mechanics, the one that enforces its entire departure from classical lines of thought'' \citep[p.~555, emphasis in the original]{Schrodinger:1935:jn}.} the feeling prevailed that entanglement was something unusual and a peculiarly microscopic phenomenon that would have to be carefully created in the laboratory (such as in an EPR-type experiment). Entanglement was regarded as an essential quantum feature that would necessarily have to be irreconcilable with classicality. These long-held beliefs likely contributed to the comparably late ``discovery'' of the theory of decoherence, which has its roots in Zeh's work from the 1970s but did not receive broader attention until Zurek's contributions appeared in the 1980s \citep{Camilleri:2009:aq}. 

It is indeed a particular irony that entanglement would turn out to be not something that had to be tamed to ensure classicality but would instead assume an important role in our understanding of aspects of the quantum-to-classical transition. Decoherence describes how entangling interactions with the environment influence the statistics of results of future measurements on the system. Decoherence provides a quantitative, dynamical explanation, wholly within standard mechanics, of the difficulty of creating and observing many superposition states and interference phenomena (most notably, those concerning ``classical'' quantities such as position), especially for mesoscopic and macroscopic systems whose many degrees of freedom typically lead to strong coupling to the environment \citep{Zeh:1970:yt,Zurek:1981:dd,Zurek:1982:tv,Zurek:2002:ii,Joos:2003:jh,Bacciagaluppi:2003:yz,Schlosshauer:2003:tv,Schlosshauer:2007:un}.  

Thus, decoherence provides a partial solution to the dynamical problem of the quantum-to-classical transition, insofar as decoherence describes effective restrictions on the superposition principle for subsystems interacting with other systems. Decoherence addresses what  \citet{Bub:2012:oo} has called a \emph{consistency problem}: the agreement between the probability distributions obtained from quantum mechanics and classical mechanics in the relevant cases. It is in this narrow qualified sense that we may say that ``classicality,'' so to speak, ``emerges'' out of the quantum formalism.

A purely dynamical account such as decoherence, however, must be viewed in the context of the series of similar attempts made by Bohr's followers (see Sec.~\ref{sec:irrev-emphdyn-probl}). Its relationship to Bohr's views regarding the object--instrument divide and the primacy of classical concepts should therefore be judged in the same way as those earlier attempts: as something that was believed by many of Bohr's followers to be in natural harmony with Bohr's views. The dynamical problem of the quantum--classical transition, as many of Bohr's followers recognized, in fact addressed a problem quite different from the chiefly epistemological problem that had preoccupied Bohr. Of course, if classical concepts are understood in a purely pragmatic sense---as something we simply \emph{do} use when we perform a measurement---decoherence may supply a justification for their use. In this particular reading of classical concepts, decoherence not only tells us \emph{why} the concepts of classical physics are applicable in the macroscopic situations relevant to our experience despite the underlying quantum-mechanical description of the world, but also \emph{when} and \emph{where} these concepts can be applied. On the other hand, insofar as decoherence is simply a consequence of a realistic application of the standard quantum formalism, which must itself be given a physical interpretation, one may suggest that the use of classical concepts has already been presupposed in any experimental setup, and that therefore any ``derivation'' of classical concepts from the quantum formalism is simply circular.

\section{Concluding remarks}

``It would be a misconception,'' Bohr wrote in 1929, ``to believe that the difficulties of atomic theory may be evaded by eventually replacing the concepts of classical physics by new conceptual forms'' \citep[p.~ 294]{Bohr:1985:mn}. Reiterating this point the same year in his ``Introductory Survey,'' he insisted that it no longer seemed likely ``that the fundamental concepts of classical theories will ever become superfluous for the description of physical experience'' \citep[p.~16]{Bohr:1987:aw}.

Such remarks have often been interpreted as indicating some kind of Kantian element in Bohr's philosophy. Yet the real problem for Bohr was not so much that we lack the cognitive faculties or the capacity to develop new concepts or a new language for describing the quantum world, but that the very success of quantum mechanics lies precisely in accounting for a vast array of experimental evidence, which itself depends on the basic conceptual framework of classical physics. As we have shown in this paper, Bohr's view of the indispensability of classical concepts is primarily epistemological in nature. To put it in Kroes's terms, Bohr was convinced that ``the structural description of physical reality rests implicitly on a functional description of at least part of the world'' \citep[p.~74]{Kroes:2003:zz}. Bohr was convinced that any functional description of an experiment in physics must be a classical description, and it was in this sense that he maintained that physics could not do without classical concepts. 

Bohr's \emph{epistemological} thesis concerning a functional description of \emph{experiment} should be distinguished from the efforts, which gathered momentum in the 1960s, to supply a \emph{dynamical} explanation of the emergence of classicality from quantum \emph{theory}. What such approaches have aimed for---and, in the case of decoherence, with great success---is to ensure, through consideration of the dynamics, a consistency between the quantum and classical descriptions (expressed in terms of the relevant probabilistic predictions) particularly for mesoscopic and macroscopic systems. While such approaches can provide, \emph{post facto}, a physical, pragmatic justification for the applicability of classical concepts, they do not touch on Bohr's main concern of an epistemology of experiment that underlies his doctrine of classical concepts. Indeed, his doctrine was \emph{premised} on the assumption that a quantum-mechanical treatment of the measurement apparatus would be ``essentially equivalent with a classical description.'' To this extent, Bohr's doctrine actually rests on the \emph{presupposition} that we can, for all intents and purposes, treat the measuring apparatus as a classical system. Indeed, it was Bohr's conviction that if this presupposition did not hold, it would be impossible to acquire knowledge about quantum objects by means of experiment. 

In this sense, one may say that Bohr's fundamental point was that any interpretation of quantum mechanics must in the end fall back on the use of classical concepts, because such concepts play an indispensible role in experimental contexts in which we acquire empirical knowledge of the world. Whether Bohr was correct in this assertion is a matter requiring further philosophical reflection, but a discussion of this point would take us beyond the scope of this paper. Our purpose here was not to present a defense of Bohr's view, but to offer a reading of Bohr---a reading that suggests that the possibility of a peaceful coexistence between Bohr's philosophy and the insights brought about by decoherence theory is more viable than has often been claimed.


\end{document}